\documentclass[aps,preprint,groupedaddress]{revtex4}
\usepackage{graphicx}
\usepackage{amssymb}
\usepackage{amsmath}
\usepackage{epstopdf}
\usepackage{longtable}

\begin{document}


\title{ Soft lubrication}
\author{J.M. Skotheim and L. Mahadevan$^*$}
\affiliation{Department of Applied Mathematics and Theoretical Physics, Centre for Mathematical Sciences, University of Cambridge, Cambridge CB3 0WA, United Kingdom}

\date{\today}

\begin{abstract}
We consider some basic principles of fluid-induced lubrication at soft interfaces. In particular, we show how  the presence of a soft substrate leads to an increase in the physical separation between surfaces sliding past each other. By considering the model problem of a symmetric non-conforming contact moving tangentially to a thin elastic layer, we determine the normal force in the small and large deflection limit, and show that there is an optimal combination of material and geometric properties which maximizes the normal force. Our results can be generalized to a variety of other geometries which show the same qualitative behavior. Thus they are relevant in the elastohydrodynamic lubrication of  soft elastic and poroelastic gels and shells, and in the context of  bio-lubrication  in cartilaginous joints.

\end{abstract}



\maketitle

Lubrication between two contacting surfaces serves to prevent adhesion and wear, and to reduce friction. The presence of an intercalating "lubricating" fluid aids both, but  gives rise to large hydrodynamic pressures in the narrow gap separating the  surfaces and can thus lead to deformations of the surfaces themselves. For stiff materials such as metals the  pressure required for noticeable deformations is very large ($\sim1$GPa) and under these conditions the lubricating fluid will exhibit non-Newtonian properties  \footnote{The  viscosity depends on the pressure and temperature \cite{Ha94}.}. However, if these surfaces are soft, as in the case of gels and thin shells, elastohydrodynamic effects can become important when the fluid is still Newtonian since the pressure required to displace the surface is appreciably less. This type of situation is also common in mammalian joints where  the synovial fluid serves as the lubricant between the soft thin cartilaginous layers which coat the much stiffer bones. Motivated by these observations, in this letter we consider the coupling between fluid flow and elastic deformation in confined geometries that are common in lubrication problems. 

As a prelude to our discussion we consider the steady motion of a cylinder of radius $R$ completely immersed in fluid and moving with a velocity $V$, with its center at height $h_0+R$ above a rigid surface (see Fig. 1). The dynamics of the fluid of viscosity $\mu$, and density $\rho$ are described using the Navier-Stokes equations 
\begin{eqnarray}
\label{1}
\rho(\partial_t {\bf v} + {\bf v}\cdot\nabla{\bf v}) = \mu\nabla^2{\bf v} - \nabla p, \\
\label{2}
\nabla\cdot{\bf v}=0,
\end{eqnarray}
where ${\bf v}$ is the 2-D velocity field $(u,w)$ and $p$ is the pressure.
Comparing the ratio between the inertial and viscous forces in the narrow gap having a 
contact length $l\sim\sqrt{Rh_0}$~\footnote{All non-degenerate (non-conforming)
contacts may be approximated as a parabola in the vicinity of the contact region, in which 
case the gap profile $h=h_0(1+\frac{x^2}{2Rh_0})$.}, we find the gap Reynolds number
Re$_{\hbox{g}} = \frac{\rho V^2/l}{\mu V/h_0^2} \sim \frac{\rho V h_0^{3/2}}{\mu R^{1/2}}
\ll \frac{\rho V R}{\mu} = $ Re, the nominal Reynolds number.  The gap Reynolds number
is small since $h_0\ll R$ and we can safely neglect the inertial terms and use the Stokes' equations 
(and the lubrication approximation thereof \cite{Ba}) to describe the hydrodynamics.
The temporal reversibility associated with the Stokes equations and the symmetry of the parabolic contact leads to the conclusion that there can be no normal force due to the horizontal motion of the cylinder. However, if there is a thin soft elastic layer on either the cylinder or the
wall, the deformation of the layer breaks the contact symmetry and leads to a normal force.
This then leads to an enhanced physical separation  and a reduced shear  so that it may be a likely cause for the low wear properties of cartilaginous joints.   

\begin{figure}
\includegraphics[width=\columnwidth]{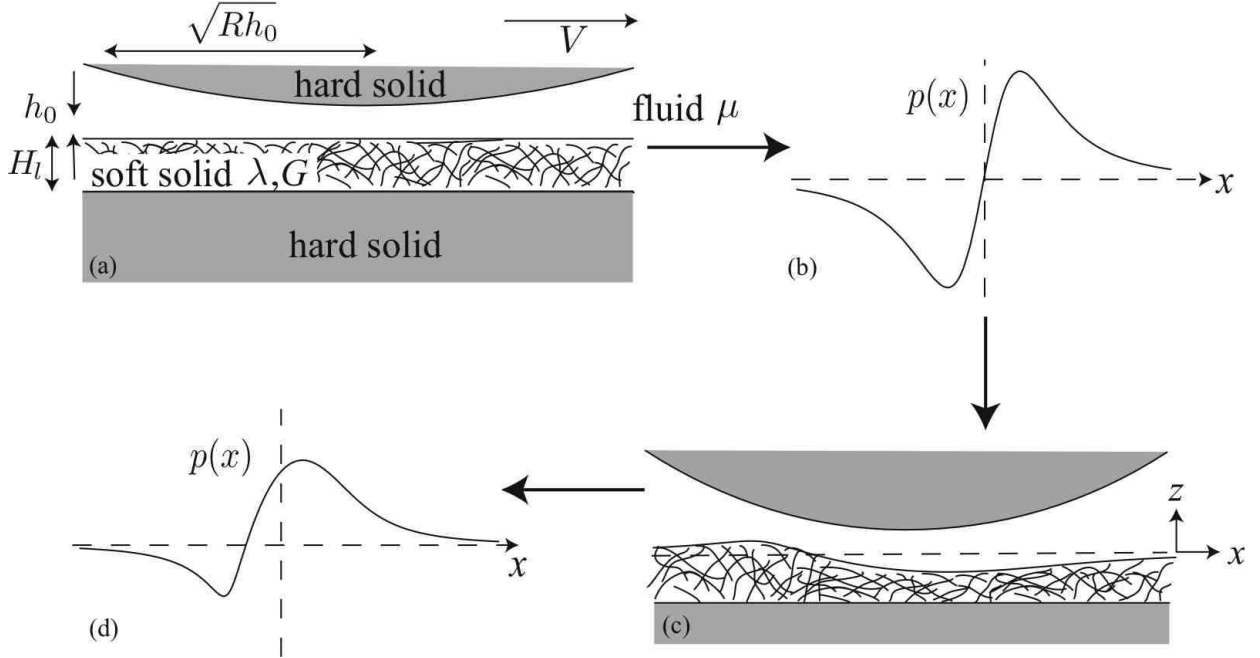}
\caption{\label{f.1}A rigid cylinder moves at a velocity $V$ a distance $h_0$ above a rigid substrate coated with an elastic layer of thickness $H_l$.  $H_l,~h_0 \ll \sqrt{h_0R}=l$.
We illustrate the steps of the perturbation analysis: (b) an antisymmetric pressure distribution
pushes down on the gel in front of and pulls the gel up behind the cylinder; (c) the fore-aft
gap profile symmetry is broken; (d) the new pressure field produces a normal force.  
(a) and (b) correspond to an undeformed substrate, while (c) and (d) correspond to solutions of (\ref{Req}), (\ref{Reqbc}) and (\ref{Hx}) for $\eta=\frac{\Delta h}{h_0}=10$.}
\end{figure}

\begin{figure}
\includegraphics[width=\columnwidth]{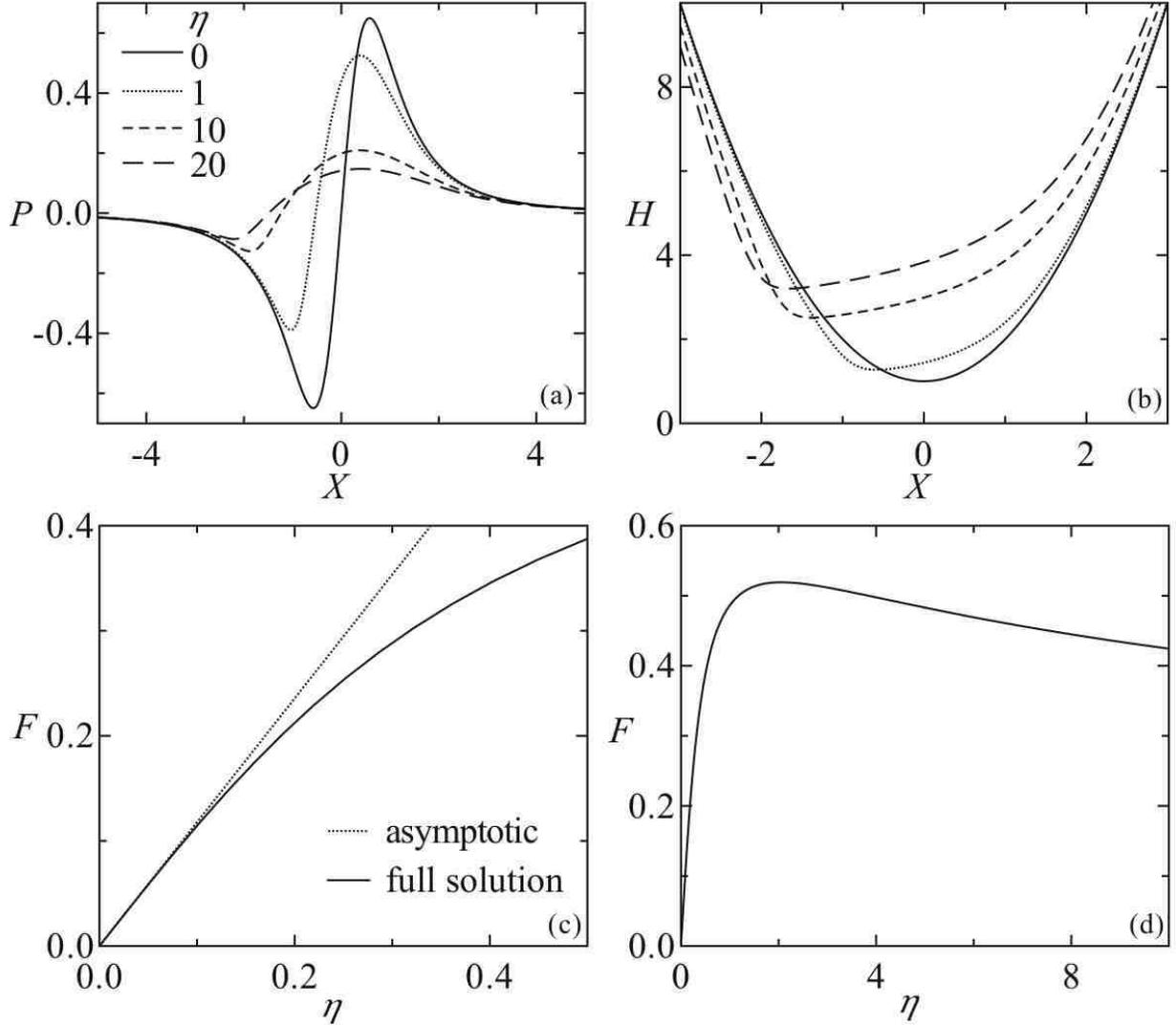}
\caption{\label{f.3}(a) The dimensionless pressure distribution, $P$, for several values of 
$\eta$, a measure of the deflection of the elastic layer compared to the
initial separation. (b) The dimensionless gap thickness profile, $H = 1 + X^2 + \eta P$.
The gap thickness and asymmetry increase with $\eta$, while the 
maximum value of the pressure decreases.
(c) For small $\eta$ asymptotic analysis predicts a dimensionless lift force 
$F=\frac{3\pi}{8}\eta$, which 
matches the numerical solution. 
(d) $F$ has a maximum at $\eta=2.06$ as a 
result of the competition between symmetry breaking and decreasing pressure.}
\end{figure}

Continuing our analysis in the context of a cylinder moving along a planar wall, we take the $x-$direction to be parallel to the wall in the direction of motion of the cylinder and the $z-$direction to be perpendicular to the wall; $p$ is the fluid pressure; $h$ is the distance between the solid surfaces. Guided by lubrication theory \cite{Ba} we use the following scalings
\begin{eqnarray}
x=\sqrt{2h_0R} X,~~~z=h_0Z,~~~p=\frac{\sqrt{2R}\mu V}{h_0^{3/2}}P, \nonumber \\
h = h_0H,~~~u=V U,
~~~w=\frac{V \sqrt{h_0}}{\sqrt{2 R}}W,
\end{eqnarray}
to reduce (\ref{1}) and (\ref{2}) to 
\begin{eqnarray}
\partial_X P=\partial_{ZZ} U, ~~
\partial_Z P = 0, \label{dimeq1}\\
\partial_X U + \partial_Z W =0. \label{dimeq2}
\end{eqnarray}
We consider steady motion in the reference frame of the cylinder, so that  the boundary conditions are
\begin{eqnarray}
U(X,0)=-1,~~U(X,H)=0,\nonumber \\ 
~W(X,0)=W(X,H)=0, \label{dimbc}
\end{eqnarray}
Integrating (\ref{dimeq1},\ref{dimeq2},\ref{dimbc}) leads to the dimensionless Reynolds equation ~\cite{Re86}:
\begin{equation}
\label{Req}
0=\partial_X(6H+H^3\partial_XP).
\end{equation}
Since the gap pressure is much larger than the ambient pressure, we may approximate the boundary conditions on the pressure field as
\begin{eqnarray}
\label{Reqbc}
P(\infty)=P(-\infty)=0.
\end{eqnarray}
Next, we consider the deformation of the elastic layer of thickness $H_l$ that rests on a rigid
support. Balance of stresses in the solid leads to 
\begin{equation}
\nabla\cdot \boldsymbol{\sigma}=0,
\end{equation}
with the stress given by
\begin{equation}
\boldsymbol{\sigma} = G (\nabla {\bf u} + \nabla{\bf u}^T) + \lambda\nabla \cdot{\bf u}\, {\bf I},
\end{equation}
where  ${\bf u}=(u_x,u_z)$ is the displacement field and $G$ and $\lambda$ are the Lam\'e constants for the solid, which is assumed to be isotropic and linearly elastic. To calculate the increase in gap thickness $H(x)$ we use the analog of the lubrication approximation in the solid layer \cite{Jo85}.  
The length scale in the $z-$direction is $H_l$ and the length scale in the $x-$direction is
$\sqrt{h_0R}$. We take the thickness of the solid layer to be small compared to the thickness of the 
contact zone, $\sqrt{h_0R} \gg H_l$, and consider a compressible elastic material, $G \sim \lambda$, 
to find the vertical force balance: 
$\partial_{zz} u_z  = 0.$
The boundary condition at the solid-fluid interface is $\boldsymbol{\sigma}\cdot{\bf n} = -p{\bf n}$, so that   $(2G+\lambda)\partial_zu_z(x,0) = -p(x)$.  Using the zero displacement condition at the interface between the soft and rigid solid, $u_z(x,-H_l)=0$ leads to the following expression for the displacement of the surface
\begin{equation}
u_z(x,0) = -\frac{H_l p(x)}{2G+\lambda}.
\end{equation}
Then the dimensionless version of the gap thickness, $h = h_0+\frac{x^2}{2R} - u_z(x,0)$, is 
\begin{equation}
 \label{Hx}
H(X)=1+X^2+\eta\,P(X),
\end{equation}
where $\eta=\Delta h/h_0 =  \frac{\sqrt{2R}H_l \mu V}{h_0^{5/2}(2G + \lambda)}$ 
is the dimensionless parameter governing the size of the deflection. Inspired by the some recent experiments \cite{MaCl02} in a similar geometry, we consider a cylinder of radius $R=10$ cm coated with a rubber layer ($H_l = 0.1$ cm, $G=1$ MPa)  moving through water ($\mu = 1$ mPa$\cdot$s, $V=1$ cm/s,  $h_0 = 10^{-3}$ cm). Then $\eta = 10^{-2}<<1$, so that we may use the perturbation expansion
$P = P_0 + \eta P_1$, where $P_0$ is the anti-symmetric pressure distribution corresponding to an undeformed layer, and $P_1$ is the  symmetric pressure perturbation induced by elastic deformation.
Substituting (\ref{Hx}) into (\ref{Req}) leads to the following equations for $P_0, P_1$:
\begin{eqnarray}
\label{5}
\eta^0 : ~\partial_X[6(1+X^2)+(1+X^2)^3\partial_XP_0]=0,\\
\label{6}
\eta^1 :~\partial_X[6 P_0 + 3 (1+X^2)^2 P_0 \partial_XP_0 + \nonumber\\
 (1+X^2)^3\partial_XP_1]=0,
\end{eqnarray}
subject to the boundary conditions $P_0(\infty)=P_0(-\infty)=P_1(\infty)=P_1(-\infty)=0$. 
Solving (\ref{5}) and (\ref{6}) yields
\begin{equation}
\label{7}
P = \frac{2 X}{(1+ X^2)^2} + \eta  \frac{3(3-5X^2)}{5(1+X^2)^5}.
\end{equation}
Then the normal force is 
\begin{equation}
F = \int_{-\infty}^\infty P\,dX =  \frac{3\pi}{8} \eta,
\end{equation}
In dimensional terms,  
$
F = \frac{3\sqrt{2}\pi}{4}\frac{\mu^2V^2H_lR^{3/2}}{h_0^{7/2}(2G+\lambda)},
$
whose scaling matches the result reported in~\cite{SeLi93}, but with a different pre-factor.
When $\eta$ is not small, we solve (\ref{Req}),(\ref{Reqbc}) and (\ref{Hx}) numerically. Figure \ref{f.3} shows that as $\eta$ increases the mean gap increases and its profile becomes  asymmetric, resembling the profile of a rigid slider bearing, a configuration well known to generate lift forces \cite{Ba}.  In addition, this increase in the gap size causes the peak pressure to decrease since $p\sim \frac{\mu V R^{1/2}}{h_0^{3/2}}$.  These two competing effects produce a maximum lift force when $\eta = 2.06$.

\begin{figure}
\includegraphics[width=\columnwidth]{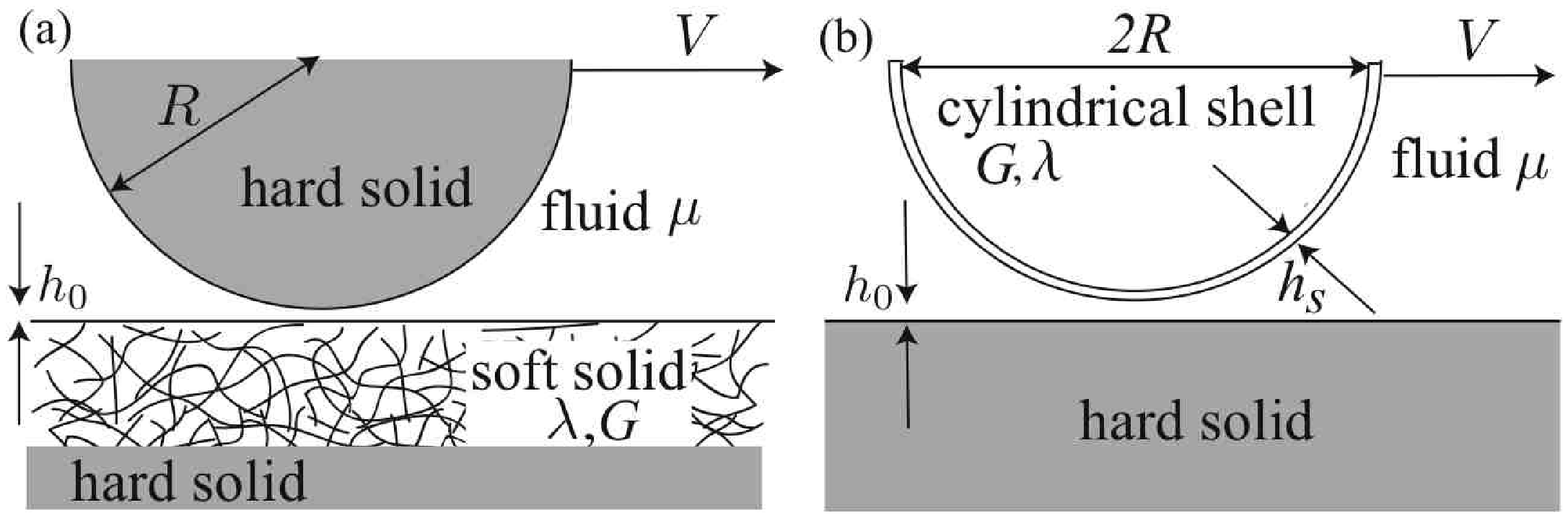}
\caption{\label{f.2}Schematic diagrams of two configurations considered on the level of scaling: (a) a soft solid coats a rigid solid where $H_l \gg \sqrt{h_0R}$, {\it i.e.} the layer thickness is larger than the 
length scale of the hydrodynamic interaction; (b) the cylinder is replaced by a cylindrical shell.}
\end{figure}

The physical basis for the previous arguments can be more easily understood using scaling and therefore allows us to generalize these results to a variety of configurations involving lubrication of soft contacts (Fig. 3; Table 1).
Balancing the pressure gradient in the gap with the viscous stresses yields 
\begin{equation}
\label{pscale}
\frac{p}{l}\sim \frac{\mu V}{h^2} \to p \sim \frac{\mu V R^{1/2}}{h^{3/2}},
\end{equation}
Substituting $h=h_0+\Delta h$, with $\Delta h \ll h_0$, we find that the lubrication pressure is
\begin{equation}
p  \sim  \frac{\mu V \sqrt{Rh_0}}{(h_0+\Delta h)^2} \sim  \frac{\mu V R^{1/2}}{h_0^{3/2}}(1 + 
 \frac{\Delta h}{h_0}) = p_0 + \frac{\Delta h}{h_0} p_1.
\end{equation}
Here $p_0$ does not contribute to the lift for the reasons outlined earlier, so that the 
the lift on the cylinder per unit length is
\begin{eqnarray} 
\label{cl}
F \sim \frac{\Delta h}{h_0} p_1 l  \sim \frac{\mu V R}{h_0^2}\Delta h,
\end{eqnarray}
where  $\Delta h$ is determined by the solution of the elasticity problem.

\begin{table*}
\caption{\label{t.1} Summary of scaling results for small surface deflections.}
\begin{ruledtabular}
\begin{tabular}{cccc}
Geometry  & Material & Surface displacement  & Lift force/unit length \\
\hline  \hline
Thin layer  & Compressible elastic solid & $\frac{ \mu V}{G} \frac{ H_l R^{1/2}}{h_0^{3/2} }$ & $\frac{\mu^2V^2}{G}\frac{H_lR^{3/2}}{h_0^{7/2}}$ \\
\hline
Thin layer & Incompressible elastic solid & $ \frac{\mu V}{G} \frac{H_l^3}{ h_0^{5/2} R^{1/2} }$& $\frac{\mu^2V^2}{G} \frac{H_l^3R^{1/2}}{h_0^{9/2}} $ \\
\hline
Thin layer & Poroelastic solid & $\frac{ \mu V}{G_{eff}} \frac{ H_l R^{1/2}}{h_0^{3/2} }$ & $\frac{\mu^2V^2}{G_{eff}}\frac{H_lR^{3/2}}{h_0^{7/2}}$ \\
\hline
Thick layer & Elastic solid & $ \frac{\mu V}{G} \frac{R}{h_0}$ & $\frac{\mu^2V^2}{G} \frac{R^2}{h_0^3} $ \\
\hline
Cylindrical Shell & Elastic solid & $ \frac{\mu V}{G} \frac{R^{7/2}}{h_s^3h_0^{1/2}}$ & $\frac{\mu^2V^2}{G} \frac{R^{9/2}}{h_s^3h_0^{5/2}}$ \\
\end{tabular}
\end{ruledtabular}
\end{table*}

For a thin compressible layer, the case treated above, the normal strain is $\frac{\Delta h}{H_l}\sim\frac{p_0}{G}
\sim \frac{\mu V R^{1/2}}{G h_0^{3/2}}$. Therefore, 
\begin{equation}
\label{17}
\Delta h \sim \frac{ \mu V}{G} \frac{ H_l R^{1/2}}{h_0^{3/2} }, ~~~F \sim  \frac{\mu^2V^2}{G}\frac{H_lR^{3/2}}{h_0^{7/2}}.
\end{equation}
In sharp contrast, a thin incompressible layer will deform via shear  with an effective shear strain
$\frac{\Delta u}{H_l}\sim \frac{l\Delta h}{H_l^2} $\footnote{An incompressible solid must satisfy the
continuity equation $\nabla\cdot{\bf u}=0$, which implies that $\frac{\Delta u}{l} \sim \frac{\Delta h}{H_l}$.}.
Balancing the elastic energy 
$\int G (\frac{R^{1/2}h_0^{1/2} \Delta h}{H_l^2})^2dV 
\sim G (\frac{R^{1/2}h_0^{1/2} \Delta h}{H_l^2})^2 H_l \sqrt{Rh_0}$ 
with the work done by the pressure $p_0 \Delta h \sqrt{Rh_0}$ 
yields $\Delta h$ in terms of $p_0$.  Then, (\ref{pscale}) and (\ref{cl}) give
\begin{equation}
\label{18}
\Delta h \sim \frac{ \mu V}{G} \frac{ H_l^3 }{h_0^{5/2}R^{1/2} }, ~~~ F\sim  \frac{\mu^2V^2}{G}\frac{H_l^3 R^{1/2}}{h_0^{9/2}}
\end{equation}

A  thick layer ($H_l\gg \sqrt{h_0R}$) may be treated as an elastic half space.  
The strain scales as $\frac{\Delta h}{\sqrt{h_0R}}$ and remains appreciable in a region of size $h_0R$.
Balancing the elastic energy $\int G (\frac{\Delta h}{\sqrt{h_0R}})^2dV \sim G (\frac{\Delta h}{\sqrt{h_0R}})^2 Rh_0$ with the work done by the pressure $p_0 \Delta h \sqrt{Rh_0}$ 
yields $\Delta h$ in terms of $p_0$. Then, (\ref{pscale}) and (\ref{cl}) give
\begin{equation}
\Delta h \sim \frac{ \mu V}{G} \frac{R}{h_0}, ~~~ F\sim  \frac{\mu^2V^2}{G}\frac{R^2}{h_0^3}
\end{equation}

Finally, we consider the case of a cylindrical shell, of radius $R$ and thickness $h_s$, 
moving over a rigid substrate.  Since the shell is thin it can be easily deformed via cylindrical bending without stretching. The bending strain is $\frac{h_s \Delta h}{R^2}$ so that the elastic energy
$\int G (\frac{h_s \Delta h}{R^2})^2 dA\sim \frac{G h_s^3 \Delta h^2}{R^3}$. Balancing this with the work done by the pressure $p_0  (\sqrt{R h_0})^2\frac{\Delta h}{R}$ yields $\Delta h$ in terms of $p_0$.  
Then, (\ref{pscale}) and (\ref{cl}) give
\begin{equation}
\Delta h \sim \frac{\mu V}{G} \frac{R^{7/2}}{h_s^3h_0^{1/2}}, ~~~ F\sim  \frac{\mu^2V^2}{G}\frac{R^{9/2}}{h_s^3h_0^{5/2}}
\end{equation}
We note that the above scalings for cylindrical contacts can be trivially generalized to spherical contacts for the case of small deformations, but space precludes us from discussing these in detail. 

We conclude with a discussion of how our results may be applied to the lubrication of cartilaginous joints~\cite{MoGu02, MoHo84}, where a thin layer of a fluid-filled gel, the cartilage, coats the stiff bones and mediates the contact between them. Here,  electrostatic effects prevent physical contact of the surfaces under high static normal loads, while elastohydrodynamic effects could enhance separation and thus reduce wear. Inspired by the treatment of cartilage using poroelasticity \cite{Gr78, MoGu02}, the continuum description of a material composed of an elastic solid  skeleton and an interstitial fluid \cite{Bi41}, we treat the cartilage layer as an isotropic poroelastic material \footnote{We will ignore screened electrostatic effects to leading order in the elastohydrodynamic problem.}.
The gel can then be described by its fluid volume fraction $\alpha \sim O(1)$, drained shear modulus $G$ and  drained bulk modulus $K\sim G$, thickness $H_l$, permeability $k$, 
and interstitial fluid viscosity $\mu$.  
Using dimensional reasoning, we can construct a poroelastic time scale
\begin{equation}
\tau_p \sim \frac{\mu H_l^2}{k K},
\end{equation}
which characterizes the time for the diffusion of stress over the layer thickness $H_l$ due to fluid flow.
Then the response of the gel is governed by the relative size of $\tau_p$ to  the time scale of the motion, $\tau \sim l/V=\sqrt{h_0R}/V$.  If $\tau \gg \tau_p$, the motion is so slow that the interstitial fluid plays no role in supporting the load.  If $\tau \sim \tau_p$, the fluid supports some of the load transiently, thereby stiffening the gel. Finally, if $\tau\ll\tau_p$ the response of the gel will depend on the size of the Stokes' length $l_s\sim \sqrt{\tau \mu/\rho}$.  If $l_s \sim H_l$,
there is no relative motion between the fluid and the solid and the gel behaves as an incompressible elastic solid~\cite{SkMa04, BuKe81}, with shear modulus $G$. From (\ref{17}) and (\ref{18}) we see that the effective modulus is
\begin{equation}
G_{eff} \sim p_0 H_l/\Delta h \sim  \frac{l^2}{H_l^2} G \sim  \frac{h_0R}{H_l^2} G.
\end{equation}
To find the scale of the deflection and lift 
force we use the same scaling analysis as for a thin compressible elastic layer
but replace $G$ with $G_{eff}(\tau)$, so that (\ref{17}) yields
\begin{equation}
\Delta h \sim \frac{ \mu V}{G_{eff}(\tau)} \frac{ H_l R^{1/2}}{h_0^{3/2} }, ~~~F \sim  \frac{\mu^2V^2}{G_{eff}(\tau)}\frac{H_lR^{3/2}}{h_0^{7/2}},
\end{equation}
where $G_{eff} \in [G, \frac{h_0R}{H_l^2} G]$. Inserting characteristic values
$V=1$ cm/sec, $G=10^7$ g/sec$^2$ cm, $H_l= 0.1$ cm, $R=1$ 
cm, $h_0=10^{-4}$ cm, $\frac{k}{\mu}=10^{-13}$ cm$^3$ sec/g
shows that $\tau \ll \tau_p$, but since $l\sim H_l$ significant material stiffening is prevented. 
Consequently, the effective modulus is $G$ and the scale of 
the deflection is 
\begin{equation}
\eta \sim \frac{\mu V}{G} \frac{ H_l R^{1/2}}{h_0^{5/2} } 
\sim 1,
\end{equation}
which suggests that joints could easily operate in a parameter regime that optimizes 
repulsive elastohydrodynamic effects. Although our estimates are based on non-conforming contact geometries; in real joints where conforming contacts are the norm, we expect a similar if not enhanced effect.

\acknowledgments
{\small We acknowledge support via the Norwegian Research Council (JS), the US Office of Naval Research Young Investigator Program (LM) and the US National Institutes of Health (LM) and the 
Schlumberger Chair Fund (L.M.).

$^*${Current address: Division of Engineering and Applied Sciences, Harvard University, 29 Oxford St., Cambridge, MA 02138.}  {\em Email  : lm@deas.harvard.edu}}

\end{document}